\documentstyle[12pt]{article}
\begin{document}

\title{\Large\bf Effective Actions for 0+1 Dimensional Scalar QED and its SUSY
Generalization at T=$\!\!\!\!$/ 0} 
\author{J. Barcelos-Neto\thanks{Permanente address: Instituto de F\'{\i}sica, 
Universidade Federal do Rio de Janeiro, RJ 21945-970, Brazil}~~and 
Ashok Das\\ 
Department of Physics and Astronomy\\
University of Rochester\\
Rochester, NY 14627 - USA}

\date{}

\maketitle

\abstract
We compute the effective actions for the 0+1 dimensional scalar field 
interacting with an Abelian gauge background, as well as for its 
supersymmetric generalization at finite temperature.

\vfill

\vspace{1cm}

\newpage

The effective action for 0+1 dimensional fermions interacting with a background
Abelian gauge field, at finite temperature, has received much attention during
the past year \cite{Du}-\cite{Barc}. In this brief report, we present the 
finite temperature effective actions for scalar fields intereracting with 
an Abelian gauge background as well as for its supersymmetric extension.

\medskip
Let us consider the interacting scalar theory described by the Lagrangian

\begin{equation}
L=\bigl(D_t\phi_j\bigr)^\ast
\bigl(D_t\phi_j\bigr)-m^2\phi_j^\ast\phi_j
\hspace{1cm}j=1,2,\dots,N_f
\label{1}
\end{equation}

\bigskip\noindent 
where

\begin{equation}
D_t\phi_j=\partial_t\phi_j+iA\phi_j
\label{2}
\end{equation}

\bigskip\noindent
Unlike the theory of massive fermions \cite{Du}, this theory is, in fact, 
invariant under charge conjugation and, consequently, the effective theory can
only contain terms involving an even number of photons. The simplest, of 
course, is the two point function and involves two distint Feynman diagrams. 
At zero temperature, this gives

\begin{eqnarray}
i\,I^{(2)}(p)&=&\int\frac{dk}{2\pi}\,
\Biggl[-\,\frac{2}{k^2-m^2+i\epsilon}
+\frac{(2k+p)^2}{\bigl(k^2-m^2+i\epsilon\bigr)
\bigl((k+p)^2-m^2+i\epsilon\bigr)}\Biggr]
\nonumber\\
&=&0
\label{3}
\end{eqnarray}

\bigskip\noindent
implying that there is no quadratic term (in $A$) in the effective action.

\medskip
As we go to higher point functions, the number of graphs increases rapidly
and a diagrammatic evaluation becomes complicated. Consequently, we follow an
alternate procedure. Let us note that the effective action obtained by
integrating out the scalar fields has the form (properly normalized)

\begin{eqnarray}
\Gamma&=&i\,N_f\log\,
\frac{\det(-D_t^2-m^2)}{\det(-\partial_t^2-m^2)}
\nonumber\\
&=&i\,N_f\,\biggl[
\log\,\frac{\det(iD_t-m)}{\det(i\partial_t-m)}
+\log\,\frac{\det(iD_t+m)}{\det(i\partial_t+m)}\biggr]
\nonumber\\
&=&\tilde\Gamma(m)+\tilde\Gamma(-m)
\label{4}
\end{eqnarray}

\bigskip\noindent 
where $\tilde\Gamma(m)$ is the effective action obtained from the much simpler,
first order scalar Lagrangian of the form

\begin{equation}
\tilde L=\tilde\phi_j^\ast\,(i\,D_t-m)\,\tilde\phi_j
\label{5}
\end{equation}

\bigskip\noindent 
Calculationally, this is much simpler. In fact, at zero temperature, this 
would lead to the same effective action as from the fermionic theory 
\cite{Du}-\cite{Barc}(except for the negative sign associated with the 
fermion loop)

\begin{equation}
\tilde\Gamma(m)=-\,\frac{N_f}{2}\,{\rm sgn}(m)\,
\int dt\,A(t)
\label{6}
\end{equation}

\bigskip\noindent
It follows from this that the effective action (\ref{4}) 

\begin{equation}
\Gamma=\tilde\Gamma(m)+\tilde\Gamma(-m)=0
\label{7}
\end{equation}

\bigskip\noindent
Namely, there is no radiative correction to the theory in Eq. (\ref{1}) at
zero temperature.

\medskip
The evaluation of the effective action, at finite temperature, also follows 
in a straightforward manner. Let us consider the theory in Eq. (\ref{5}) 
(with $m>0$ for simplicity) and note that because of the bosonic nature of the 
fields, the propagator at finite temperature has the form \cite{Das2}

\begin{equation}
\tilde G(p)=\frac{i}{p-m+i\epsilon}
+2\pi\,n_B(m)\,\delta(p-m)
\label{8}
\end{equation}

\bigskip\noindent
with $(\beta=1/kT)$

\begin{equation}
n_B(m)=\frac{1}{\strut\displaystyle{\rm e}^{\beta m}-1}
\label{9}
\end{equation}

\bigskip\noindent
The propagator in coordinate space has the form

\begin{equation}
\tilde G(t)=\Bigl(\theta(t)+n_B(m)\Bigr)\,
{\rm e}^{-\,imt}
\label{10}
\end{equation}

\bigskip\noindent
The 1-pt function is easy to evaluate.

\begin{eqnarray}
i\,\tilde I^{(1)}(t)&=&-\,iN_f\,
\Bigl(\theta(0)+n_B(m)\Bigr)
\nonumber\\
&=&-\,\frac{iN_f}{2}\,\coth\,\frac{\beta m}{2}
\label{11}
\end{eqnarray}

\bigskip\noindent
This, of course, reduces to the 1-pt function from  Eq. (\ref{6}) at zero 
temperature (for $m>0$). Furthermore, following \cite{Das1}, we can show that 
in this case, the recursion relation between amplitudes has exactly the same 
form as in the fermionic theory

\begin{equation}
\frac{\partial\tilde I^{(N)}}{\partial m}
=-\,i\beta\,(N+1)\,\tilde I^{(N+1)}
\label{12}
\end{equation}

\bigskip\noindent
so that all the amplitudes are related to the 1-pt amplitude recursively. 
Thus, in this case, the effective action takes the form $(a=\int dt A(t))$

\begin{eqnarray}
\tilde\Gamma(m)&=&-\,i\sum_{N=1}^\infty a^N\,
(i\tilde I^{(N)})
\nonumber\\
&=&-\,\frac{i\beta N_f}{2}\,\sum_{N=1}^\infty
\frac{1}{N!}\,\biggl(\frac{ia}{\beta}\biggr)^N\,
\frac{\partial^{N-1}}{\partial m^{N-1}}\,\coth\,
\frac{\beta m}{2}
\nonumber\\
&=&i\,N_f\,\log\,\Bigl\{\cos\,\frac{a}{2}
+i\,\coth\,\frac{\beta m}{2}\,\sin\,\frac{a}{2}\Bigr\}
\label{13}
\end{eqnarray}

\bigskip\noindent
Consequently, the effective action for the inteacting scalar theory in 
Eq. (\ref{1}) has the form (see Eq. (\ref{4}))

\begin{eqnarray}
\Gamma&=&\tilde\Gamma(m)+\tilde\Gamma(-m)
\nonumber\\
&=&i\,N_f\,\log\,\Bigl\{\cos^2\frac{a}{2}
+\coth^2\,\frac{\beta m}{2}\,\sin^2\,\frac{a}{2}\Bigr\}
\label{14}
\end{eqnarray}

\bigskip\noindent
This can be easily seen to reduce to the zero temperature result of Eq. 
(\ref{7}) and is invariant under the large gauge transformation 
$a\longrightarrow a+2\pi\,N$ as well, for any number of flavors, $N_f$.

\bigskip
Let us next consider the supersymmetric generalization of Eq.(\ref{1}),

\begin{equation}
L_{\rm SUSY}=(D_t\phi_j)^\ast(D_t\phi_j)
-m^2\,\phi_j^\ast\phi_j
+\bar\psi_j(iD_t-m)\psi_j
\label{15}
\end{equation}

\bigskip\noindent
The supersymmetric multiplet, $(\phi_j,\psi_j)$, is interacting with a 
background, Abelian gauge field. There is no photino in this theory and yet, 
because of the simplicity of 0+1 dimension, it can be easily checked that 
this Lagrangian is invariant under the supersymmetry transformations

\begin{eqnarray}
\delta\,\phi_j&=&\epsilon\,\psi_j
\nonumber\\
\delta\,\psi_j&=&-\,i(D_t\phi_j)\,\epsilon
\nonumber\\
\delta\,\lambda&=&A\,\epsilon
\nonumber\\
\delta\,A&=&-\,i\epsilon\,\partial_t\lambda
\label{16}
\end{eqnarray}

\bigskip\noindent
where we assume that a Majorana fermion is a real fermion, $\epsilon^\ast=
\epsilon$ and $\bar\psi=\psi^\ast$. The transformations of the scalar and the 
charged fermion, in (\ref{16}), are the conventional ones whereas those of 
the photon and the photino look unconventional. This is primarily because in 
0+1 dimension, the photon has no kinetic energy term and is like an 
auxiliary field and that a fermion with an auxiliary field can describe a 
supersymmetric multiplet in lower dimensions \cite{Sa,Am} (the conventional 
SUSY transformation would imply that the photino does not transform since 
the field strengh associated with the gauge field vanishes). In fact, it is 
worth noting that a Chern-Simons term in 0+1 dimension would be automatically 
supersymmetric under such a transformation and that the transformations of 
$(A,\lambda)$ satisfy the supersymmetry algebra. This is, of course, not the 
most general supersymmetric interacting theory involving 
$\phi_j,\,\psi_j,\,A,\,\lambda$. For example, it is easy to check that 

\begin{eqnarray}
L&=& (D_t\phi_j)^\ast(D_t\phi_j)-m^2\phi^\ast_j\phi_j
+\bar\psi_j(iD_t-m)\psi_j+i\lambda\dot\lambda
\nonumber\\
&&\phantom{(D_t\phi_j)^\ast(D_t\phi_j)}
+\lambda\,(\psi_j\phi^\ast_j-\bar\psi_j\phi_j)
-\frac{1}{4}\,\Bigl(\phi_j^\ast\phi_j\Bigr)^2
\label{17}
\end{eqnarray}

\bigskip\noindent
is also invariant under the transformations (\ref{16}) on-shell. It is, 
however, not clear if this theory is soluble because of the quartic scalar 
interactions. On the other hand, it is interesting that the simple theory 
in (ref{15}) is already invariant under supersymmetry and is soluble.

\medskip
The effective action for Eq. (\ref{15}), of course, follows from 
Eq. (\ref{14}) as well as from earlier work \cite{Du} and has the form

\begin{eqnarray}
\Gamma_{\rm SUSY}&=&i\,N_f\,\biggl[\log\Bigl\{
\cos^2\,\frac{a}{2}+\coth^2\,\frac{\beta m}{2}\,
\sin^2\,\frac{a}{2}\Bigr\}
\nonumber\\
&&-\log\Bigl\{\cos\,\frac{a}{2}
+i\,\tanh\,\frac{\beta m}{2}\,\sin\,\frac{a}{2}\Bigr\}\bigg]
\label{18}
\end{eqnarray}

\bigskip\noindent
Furthermore, this action, in addition to being invariant under large gauge 
transformations, is also invariant under the supersymmetry transformations,
Eq. (\ref{16}). In fact, it is interesting to note that a SUSY 
transformation that leaves the Chern-Simons action invariant would also 
automatically be a symmetry of any nonextensive function of it. This behavior 
is quite distinct from the conventional expectation \cite{Das3} that 
supersymmetry is broken at finite temperature.

\bigskip
\noindent {\bf Acknowledgment:} This work was supported in part by U.S. DOE 
Grant No. DE-FG-02-91 ER40685, NSF-INT-9602559 and CNPq (brazilian agency).

\end{document}